\documentclass[11pt,a4paper]{article}

\usepackage[margin=35mm]{geometry}
\usepackage{macros}
\usepackage{lstcoq}
\usepackage{lstocaml}
\usepackage{amsfonts,stmaryrd}
\usepackage{amsmath}
\usepackage{graphics}
\usepackage{array}
\usepackage{paralist}
\usepackage[all]{xy}

\usepackage{caption}
\usepackage{subcaption}
\usepackage{multicol}

\usepackage{tikz}
\usetikzlibrary{trees}
\usepackage{microtype}

\usepackage[backend=bibtex]{biblatex}
\usepackage{dmbiblatex}
\usepackage{hyperref}
\hypersetup{pdftitle={Implementing hash-consed structures in Coq},
  pdfauthor={Thomas Braibant, Jacques-Henri Jourdan and David Monniaux}}

\author{Thomas Braibant\thanks{\href{http://www.inria.fr/}{INRIA}} \and Jacques-Henri Jourdan\footnotemark[2] \and David Monniaux\thanks{\href{http://www.cnrs.fr/}{CNRS} / \href{http://www-verimag.imag.fr/}{VERIMAG}}}
\title{Implementing hash-consed structures in Coq\thanks{This work was partially funded by ANR project ``\href{http://verasco.imag.fr/}{VERASCO}'' and ERC project ``\href{http://stator.imag.fr/}{STATOR}''}}

\bibliography{bib}

\begin{document}
\maketitle

\begin{abstract}
  We report on three different approaches to use hash-consing in
  programs certified with the Coq system, using binary decision
  diagrams (BDD) as running example. The use cases include execution
  inside Coq, or execution of the extracted OCaml code. There are
  different trade-offs between faithful use of pristine extracted
  code, and code that is fine-tuned to make use of OCaml programming
  constructs not available in Coq. We discuss the possible
  consequences in terms of performances and guarantees.
\end{abstract}

\section{Introduction}
\emph{Hash-consing} is an implementation technique for immutable data
structures that keeps a single copy, in a global hash table, of
semantically equivalent objects, giving them unique identifiers and
enabling constant time equality testing and efficient
\emph{memoization} (also known as \emph{dynamic programming}).
A prime example of the use of hash-consing is reduced ordered binary
decision diagrams (ROBDDs, BDDs for short), representations of
Boolean functions~\cite{TAOCP_BDD} often used in software and hardware
formal verification tools, in particular \emph{model checkers}.

A Boolean function $f: \{0,1\}^n \rightarrow \{0,1\}$ can be
represented as a complete binary tree with $2^n-1$ decision nodes,
labeled by variables $x_i$ according to the depth from the root (thus
the adjective \emph{ordered}) and with subtrees labeled $0$ and $1$,
and leaves labeled \coqe{T} (for true) or \coqe{F} (for false) .
Such a tree can be \emph{reduced} by merging identical subtrees, thus
becoming a connected \emph{directed acyclic graph} (see second diagram
below); choice nodes with identical children are removed
(see third diagram below).
The reduced representation is \emph{canonical}: a function is (up to
variable ordering $x_1,\dots,x_n$) represented by a unique ROBDD.

For instance, the function $f(0,0)=\texttt{T}$, $f(0,1)=\texttt{F}$, $f(1,0)=\texttt{T}$,
$f(1,1)=\texttt{F}$ is represented, then simplified as:

\smallskip

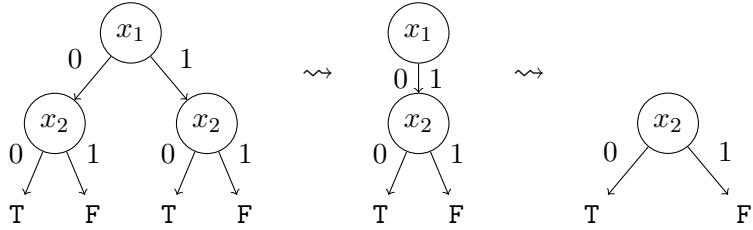
\begin{center}
\begin{tikzpicture}[level/.style={level distance=12mm,sibling distance=20mm/#1}]
\node[circle,draw] (t) {$x_1$}
  child {
    node[circle,draw] (t0) {$x_2$}
      child { node (t00) {\texttt{T}} edge from parent[->] node[above left] { $0$ } }
      child { node (t01) {\texttt{F}} edge from parent[->] node[above right] { $1$ } }
    edge from parent[->] node[above left] { $0$ }
  }
  child {
    node[circle,draw] (t1) {$x_2$}
      child { node (t10) {\texttt{T}} edge from parent[->] node[above left] { $0$ } }
      child { node (t11) {\texttt{F}} edge from parent[->] node[above right] { $1$ } }
    edge from parent[->] node[above right] { $1$ }
  }
  ;
\end{tikzpicture}
\quad\raisebox{2cm}{\large$\leadsto$}\quad
\begin{tikzpicture}[level/.style={level distance=12mm,sibling distance=20mm/#1}]
\node[circle,draw] (t) {$x_1$}
  child {
    node[circle,draw] (t0) {$x_2$}
      child { node (t00) {\texttt{T}} edge from parent[->] node[above left] { $0$ } }
      child { node (t01) {\texttt{F}} edge from parent[->] node[above right] { $1$ } }
    edge from parent[->] node[left] { $0$ } node[right] { $1$ }
  }
  ;
\end{tikzpicture}
\quad\raisebox{2cm}{\large$\leadsto$}\quad
\begin{tikzpicture}[level/.style={level distance=12mm,sibling distance=20mm/#1}]
\node[circle,draw] (t0) {$x_2$}
      child { node (t00) {\texttt{T}} edge from parent[->] node[above left] { $0$ } }
      child { node (t01) {\texttt{F}} edge from parent[->] node[above right] { $1$ } };
\end{tikzpicture}
\end{center}

In practice, one directly constructs the reduced tree.  To do so, a
BDD library usually maintains a global pool of diagrams and never
recreates a diagram that is isomorphic to one already in memory,
instead reusing the one already present. In typical implementations,
this pool is a global hash table. Hence, the phrase \emph{hash
  consing} denotes the technique of replacing nodes creation by lookup
in a hash table returning a preexisting object, or creation of the
object followed by insertion into the table if previously nonexistent.
A unique identifier is given to each object, allowing fast hashing and
comparisons.
This makes it possible to do efficient \emph{memoization}: the results
of an operation are tabulated so as to be returned immediately when
an identical sub-problem is encountered. 
For instance, in a BDD library, memoization is crucial to implement
the or/and/xor operations with time complexity in $O(|a|.|b|)$ where
$|a|$ and $|b|$ are the sizes of the inputs; in contrast, the naive
approach yields exponential complexity.


In this article, we investigate how hash-consing and memoization,
imperative techniques, may be implemented using the Coq proof
assistant, using the example of a BDD library, with two possible uses:
\begin{inparaenum}[1)]
\item to be executed inside {Coq} with reasonable
  efficiency, e.g. for proofs by reflection;
\item or to be executed efficiently when extracted to {OCaml},
  e.g. for use in a model-checking or static analysis tool
  proved correct in Coq.
\end{inparaenum}

\section{A Problem and Three Solutions}
In the following, we propose to implement a BDD library using three
different approaches. We focus on a minimal set of operations: node
creation, Boolean operations (or, and, xor, not) and equality testing
on formulas represented as ROBDDs; and we provide formal guaranties of
their correctness. 
(Note that, in some of our solutions, we do not prove the completeness
of the equality test. That is,
we prove that the equality test returning true implies equality of the
formulas; but proving the converse is not essential for many
applications.)

The typical way of implementing hash-consing (a global hash table)
does not translate easily to Coq.
The reason is that the Gallina programming language at the heart
of the Coq proof assistant is a purely applicative language, without
imperative traits such as hash tables, or pointers or pointers
equality.
%
%

Therefore, there are two approaches to the implementation of
hash-consing for data-structures in Coq. 
The first one is to model the memory using finite maps inside Coq, and
use indices in the maps as surrogates for pointers, implementing all
the aforementioned operations on these persistent maps.  Such an
implementation was described in \cite{DBLP:conf/asian/VermaGPA00,verma:inria-00072797}, and
we propose a new one in \secref{sec:pure-coq}.
The second one is to recover imperative features by
fine-tuning the extraction of Coq code: either by realizing
selected Coq constants by efficient OCaml definitions, e.g.,
extracting Coq constructors into smart OCaml constructors and fixpoint
combinators into memoizing fixpoint combinators
(see~\secref{sec:smart-cons}); or by explicitly declaring as axioms
the OCaml code implementing the hash constructs and its properties
(see \secref{sec:axioms}).

\subsection{Pure Coq}\label{sec:pure-coq}

Our first implementation of BDDs is defined as follows in Coq. 
First, we assign a unique identifier to each decision node.
Second, we represent the directed acyclic graph underlying a BDD as a
Coq finite map from identifiers to decision nodes (that is, tuples
that hold the left child, the node variable and the right child).
For instance, the following graph, on the left, can be represented using
the map on the right.

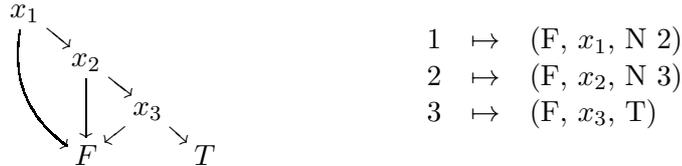
\begin{tabular}{cc}
  \begin{minipage}{0.5\linewidth}
    \begin{displaymath}
      \SelectTips{eu}{10}
      \xymatrix@R=0.5pc @C=0.5pc{
        x_1 \ar@{->}[dr] \ar@/_1pc/@{->}[dddr] & & & \\
        & x_2 \ar@{->}[dd] \ar@{->}[dr] & & \\
        & & x_3 \ar@{->}[dl] \ar@{->}[dr] & \\
        & F &  & T  
      }
    \end{displaymath}    
  \end{minipage}
  &
  \begin{tabular}{c@{$\quad\mapsto\quad$}l}
    1 & (F, $x_1$, N 2) \\
    2 & (F, $x_2$, N 3) \\
    3 & (F, $x_3$, T)
  \end{tabular}
\end{tabular}

\noindent Then, we implement the hash-consing pool using another map
from decision nodes to node identifiers and a \coqe{next} counter that
is used to assign a unique identifier to a fresh node. Equality
between BDDs is then provided by decidable equality over node
identifiers.
We present on Fig.~\ref{fig:pure-hashcons} our inductive definitions
(left) and the code of the associated allocation function
\coqe{mk_node} (right), knowing that \mbox{\coqe{upd n st}} allocates
the fresh node \coqe{n} in the hash-consing state \coqe{st} (taking
care of updating both finite maps and incrementing the ``next fresh''
counter).

\newcommand\fmap{\leadsto}

\begin{figure}[t]
  \centering
\begin{twolistings}
\begin{coq}
Inductive expr := 
  F | T | N : positive -> expr. 
Definition node := (expr * var * expr). 
Record hashcons := {
graph: positive $\fmap$ node;
hmap : node $\fmap$ positive; 
next : positive }.     
\end{coq}
&
\begin{coq}
Definition mk_node (l : expr) (v: var) (h : expr) st :=
if expr_eqb l h then (l,st)
else match find (l,v,h) (hmap st) with 
       | Some x => (N x, st)
       | None => (N st.(next), upd (l,v,h) st)
     end. 
$ $
\end{coq}
\end{twolistings} 
  \caption{Hash-consing in pure Coq}\label{fig:pure-hashcons}
\end{figure}

We define well-formedness as follows.
A node identifier is \emph{valid} in a given global state when it is lower
than the value of the \coqe{next} counter.
Then, the notion of well-formedness of global states covers the facts
that
\coqe{graph} maps all valid node identifiers to valid nodes (nodes
whose children are valid); %
and \coqe{hmap} is a left-inverse of \coqe{graph}.

Then, all operations thread the current global state in a monadic
fashion that is, of course, reminiscent of a state monad.
The correctness of BDD operations corresponds to the facts that 
\begin{inparaenum}[1)]
\item the global state is used in a monotonic fashion (that is the
structure of the resulting global state is a refinement of the input
one and that the denotation of expressions is preserved);
\item the resulting global state is well-formed;
\item the denotation of the resulting BDD expression is correct.
\end{inparaenum}
As can be expected from our data structure, BDD operations cannot be
defined using structural recursion (there is no inductive structure on
which to recurse).
Using well-founded recursion is difficult here because the
well-founded relation involves both parameters of the function and the
global state. Proving it to be well-founded would involve merging
non-trivial proofs of monotonicity within programs.
In the end, we resorted to define partial functions that use a
\emph{fuel} argument to ensure termination. 
 
\smallskip

Finally, it is possible to enrich our hash-consing structure with
memoization tables in order to tabulate the results of BDD operations.

\begin{twolistings}
\begin{coq}
Record memo := {
mand : (positive * positive) $\fmap$ expr;
mor  : (positive * positive) $\fmap$ expr;
mxor : (positive * positive) $\fmap$ expr;
mneg : positive $\fmap$ expr}.
\end{coq}
&
  \begin{coq}
Record BDD := { ... :> hashcons; ... :> memo}.



$ $   
\end{coq}
\end{twolistings}
The memoization tables are passed around by the state monad, just as
the hash-consing structure. It is then necessary to maintain
invariants on the memoization information. Namely, we have to prove
that the nodes referenced in the domain and in the codomain of those
tables are valid; and that the memoization information is semantically
correct.

As a final note: this implementation currently lacks garbage
collection (allocated nodes are never destroyed until the allocation
map becomes unreachable as a whole); it could be added e.g. by
reference counting.

\subsection{Smart constructors}\label{sec:smart-cons}

In the previous approach, we use a state monad to store
information about hash-consing and memoization. However, one can see
that, even if these programming constructs use a mutable state, they
behave transparently with respect to the pure Coq definitions.
If we abandon efficient executability inside Coq, we can write the BDD
library in Coq as if manipulating decision trees without sharing, then
add the hash-consing and memoization code by tweaking the extraction
mechanism.  An additional benefit is that, since we use native hash
tables, we may as well use \emph{weak} ones, enabling the native
garbage collector to reclaim unused nodes without being prevented from
doing so by the pointer from the table.

More precisely, we define our BDDs as in
Fig.~\ref{fig:coq-smart-hashcons}. Moreover, we tell Coq to extract
the \coqe{bdd} inductive type to a custom \ocamle{bdd} OCaml type (see
left of Fig.~\ref{fig:ocaml-smart-hashcons}) and to extract
constructors into smart constructors maintaining the maximum sharing
property. These smart constructors make use of generic hash-consing
library by Conchon and Filli\^atre~\cite{ConchonFilliatre06wml} that
defines the %
\ocamle{'a hash_consed} type of hash-consed values of type \ocamle{'a}
and the \ocamle{hashcons} function that returns a unique hash-consed
representative for the parameter.
Internally, the library uses suitable hash and equality functions on
BDDs together with weak hash tables to keep track of unique
representatives.

\begin{figure}[t]

\begin{subfigure}[t]{\textwidth}
  \centering
\begin{twolistings}
\begin{coq}
Inductive bdd: Type :=
| T | F | N: var -> bdd -> bdd -> bdd.
\end{coq}
&
\begin{coq}
Extract Inductive bdd => 
  "bdd hash_consed" ["hT" "hF" "hN"] "bdd_match".
\end{coq}
\end{twolistings} 
  \caption{BDDs in Coq as decision trees}\label{fig:coq-smart-hashcons}
\end{subfigure}

\begin{subfigure}[t]{\textwidth}
  \centering
\begin{twolistings}
\begin{ocaml}
type bdd = 
| T | F
| N of var * bdd hash_consed * bdd hash_consed

let bdd_tbl = hashcons_create 257


$ $
\end{ocaml}
&
\begin{ocaml}
let hT = hashcons bdd_tbl T
let hF = hashcons bdd_tbl F
let hN (p, b1, b2) = hashcons bdd_tbl (N(p, b1, b2))

let bdd_match fT fF fN b =
  match b.node with
  | T -> fT () | F -> fF ()
  | N(p, b1, b2) -> fN p b1 b2
\end{ocaml}
\end{twolistings} 
  \caption{Hash-consed OCaml BDD type}\label{fig:ocaml-smart-hashcons}
\end{subfigure}

\begin{subfigure}[t]{\textwidth}
  \centering
\begin{twolistings}
\begin{coq}
Definition memoFix1 :=
  Fix (well_founded_ltof bdd bdd_size).
Lemma memoFix1_eq : forall Ty F b,
  memoFix1 Ty F b =
  F b (fun b' _ => memoFix1 Ty F b').
Proof. [...] Qed.
\end{coq}
&
\begin{coq}
Program Definition bdd_not : bdd -> bdd :=
  memoFix1 _ (fun b rec => match b with
                  | T => F | F => T
                  | N v bt bf =>
                    N v (rec bt _) (rec bf _)
                end).
\end{coq}
\end{twolistings} 
  \caption{Using a fixpoint combinator for \coqe{bdd_not}}\label{fig:smart-not}
\end{subfigure}
\caption{Implementing BDDs in Coq, extracting them using smart constructors}
\end{figure}

In Coq, we define the obvious \coqe{bdd_eqb} function of type
\mbox{\coqe{bdd -> bdd -> bool},} that decides structural equality of
BDDs. 
Then, we extract this function into OCaml's physical equality. 
From a meta-level perspective, the two are equivalent thanks to the
physical unicity of hash-consed structures.

The last ingredient needed to transform a decision tree library into a
BDD library is memoization. We implement it by using special
well-founded fixpoint combinators in Coq definitions, which we extract
into a memoizing fixpoint combinator in OCaml.
As an example, we give the definition of the \coqe{bdd_not} operation
in Fig.~\ref{fig:smart-not}. The fixpoint combinator is defined using
the Coq general \coqe{Fix} well-founded fixpoint combinator that
respects a fixpoint equality property. 
The definition of \coqe{bdd_not} then uses \coqe{memoFix1} and
requires proving that the BDDs sizes are decreasing (these
trivial proof obligations are automatically discharged).

We extract the \coqe{memoFix1} combinator to a memoizing construct,
that is observationally equivalent to the original one. However, this
new construct tabulates results in order to avoid unnecessary
recursive calls.
We use similar techniques for binary operations. As all Coq
definitions are kept simple, proofs are straightforward: we can prove
semantic correctness of all operations directly using structural
induction on decision trees.

\subsection{Axioms}\label{sec:axioms}
\begin{figure}[t]
\begin{multicols}{2}
\begin{lstlisting}[language=Coq]
Axiom var : Set.

Axiom uid : Set. 
Axiom uid_eqb : uid -> uid -> bool.
Axiom uid_eq_correct: forall x y : uid,
    (uid_eqb x y =true) <-> x=y.

Inductive bdd : Set :=
| T | F 
| N : uid -> var -> bdd -> bdd -> bdd.

Axiom mkN : var -> bdd -> bdd -> bdd.

Axiom mkN_ok :
forall v : var, forall bt bf : bdd,
  exists id, mkN v bt bf = N id v bt bf.

Inductive valid : bdd -> Prop :=
| valid_T : valid T
| valid_F : valid F
| valid_N : forall var bt bf,
  (valid bt) -> (valid bf) ->
  (valid (mkN var bt bf)).

Axiom shallow_equal_ok :
  forall id1 id2 : uid,
  forall var1 var2 : var,	
  forall bt1 bf1 bt2 bf2 : bdd,
    valid (N id1 var1 bt1 bf1) ->
    valid (N id2 var2 bt2 bf2) ->
    id1 = id2 ->
    N id1 var1 bt1 bf1 =
    N id2 var2 bt2 bf2.
    
\end{lstlisting}
\end{multicols}
\caption{Axiomatization of equality using unique identifiers}\label{fig:axioms}
\end{figure}

In the previous approach, hash-consing and memoization are done after
the fact, and are completely transparent for the user. In the
following, we make more explicit the hypotheses that we make
on the representation of BDDs.
That is, we make visible in the inductive type of BDDs that each BDD
node has a ``unique identifier'' field (see Fig.~\ref{fig:axioms}) and
we take the node construction function as an axiom, which is
implemented in OCaml.
Note that nothing prevents the \Coq{} program from creating new BDD
nodes without calling this function \coqe{mkN}. Yet, only objects
created by it (or copies thereof) satisfy the \coqe{valid} predicate;
we must declare another axiom stating that unique identifier equality
is equivalent to Coq's Leibniz equality \emph{for valid nodes}.
Then, we can use unique identifiers to check for equality.

This approach is close to the previous one. It has one advantage, the
fact that unique identifiers are accessible from the Coq code. They can
for instance  be used for building maps from BDDs to other data, as
needed in order to print BDDs as a linear sequence of definitions with
back-references to shared nodes. Yet, one could also expose unique
identifiers in the ``smart constructor'' approach by stating as axioms
that there exists an injection from the BDD nodes to a totally ordered
type of unique identifiers.

The use of axioms is debatable. On the one hand, the use of axioms somewhat
lowers the confidence we can give in the proofs, and they make the code
not executable within Coq. On the other hand, these axioms are actually
used implicitly when extracting Coq constructors to ``smart constructors'':
they correspond to the metatheoretical statement that these constructors
behave as native Coq constructors. Thus, they make explicit some of the
magic done during extraction.

\section{Discussion}
We compare our approaches on different aspects:
\begin{description}
\item[Executability inside Coq.] Both the ``smart constructors'' and
  the ``pure'' implementations can be executed inside Coq, even if the
  former has dreadful performances (when executed inside Coq, it uses
  binary decision trees). The ``axiomatic'' approach cannot be
  executed inside Coq.
\item[Efficiency of the extracted OCaml code.] We have yet to perform
  extensive testing, but preliminary benchmarks indicate that the
  ``pure'' approach yields code that is roughly five times slower than
  the ``smart constructors'' approach (and we assume that the latter
  is also representative of the ``axiomatic'' approach) on classic
  examples taken from previous BDD experiments in
  Coq~\cite{DBLP:conf/asian/VermaGPA00}.
  We have yet to measure memory consumption. 
\item[Trust in the extracted code.] Unsurprisingly, the ``smart
  constructors'' and the ``axiomatic'' approaches yield code that is
  harder to trust, while the ``pure'' approach leaves the extracted
  code pristine.
\item[Proof.] From a proof-effort perspective, the ``smart
  constructors'' is by far the simplest. 
  The ``axiomatic'' approach involves the burden of dealing with
  axioms. However, it makes it easier to trust that what is formally
  proven corresponds to the real behavior of the underlying runtime.
  By comparison, the ``pure'' approach required considerably more
  proof-engineering in order to check the validity of invariants on
  the global state.
\item[Garbage collection.] Implementing (and proving correct) garbage
  collection for the ``pure'' approach would require a substantial
  amount of work. By contrast, the ``smart'' and ``axioms'' approaches
  make it possible to use OCaml's garbage collector to reclaim
  unreachable nodes.
\end{description}

\section{Conclusion and directions for future works}

In this paper, we proposed two solutions to implement hash-consing in
programs certified with the Coq system. The first one is to implement
it using Coq data-structures; the second is to use the imperative
features provided by OCaml through the tuning of the extraction
mechanism.
The difference in flavor between the mapping of Coq constants to smart
OCaml realizers or the axiomatization of there realizers in Coq is a
matter of taste. In both cases, some meta-theoretical reasoning is
required and requires to ``sweep something under the rug''. 

We conclude with directions for future works. 
First, we believe that the smart constructors approach is
generalizable to a huge variety of inductive types. One can imagine
that it could be part of the job of Coq's extraction mechanism to
implement on-demand such smart constructors and memoizers as it was
the case for other imperative
constructs~\cite{DBLP:conf/itp/ArmandGST10}.
Second, we look forward to investigate to what extent one could
provide a certified version of the hash-consing library proposed by
Conchon and Filli\^atre~\cite{ConchonFilliatre06wml}.
%

\paragraph{Ackowledgements.} 
We thank the reviewers for their helpful comments and Jean-Christophe
Filli\^atre for fruitful discussions.

\printbibliography
\end{document}